\documentclass{article}
\usepackage[preprint]{neurips_2026}
\workshoptitle{Who Verifies the Agents? Toward Reliable Agent Development}

\usepackage[utf8]{inputenc}
\usepackage[T1]{fontenc}
\usepackage{hyperref}
\usepackage{url}
\usepackage{booktabs}
\usepackage{amsfonts}
\usepackage{amsmath}
\usepackage{nicefrac}
\usepackage{microtype}
\usepackage{xcolor}
\usepackage{graphicx}
\usepackage{listings}
\usepackage{enumitem}
\usepackage{pifont}
\usepackage{multirow}
\usepackage{tabularx}

\definecolor{linkblue}{RGB}{0,51,153}
\hypersetup{colorlinks=true,linkcolor=linkblue,citecolor=linkblue,urlcolor=linkblue,pdfauthor={Xin Xu and Siru Tao},pdftitle={Discriminating Fixture Coverage in Agent-Infrastructure Verification Suites}}

\title{Discriminating Fixture Coverage in Agent-Infrastructure Verification Suites}

\author{
  Xin Xu\\
  Carnegie Mellon University\\
  Eragon\\
  \texttt{xuxin@cmu.edu}
  \And
  Siru Tao\\
  Carnegie Mellon University\\
  \texttt{sirutao@andrew.cmu.edu}
}

\newcommand{\refadapter}{\textsc{reference}}
\newcommand{\naiveadapter}{\textsc{naive}}
\newcommand{\partialadapter}{\textsc{ordering-defect}}
\newcommand{\dupdefect}{\textsc{dedup-defect}}
\newcommand{\scopedefect}{\textsc{isolation-defect}}
\newcommand{\snapdefect}{\textsc{convergence-defect}}
\newcommand{\regressdefect}{\textsc{regression-defect}}

\begin{document}

\maketitle

\begin{abstract}
Invariant suites and runtime monitors increasingly gate agent deployment decisions, and the evidence offered for any particular suite is almost always a single observation: it passes an implementation believed correct and fails one believed broken. We measure what that observation is worth. Applying mutation analysis to an invariant suite for a multi-session agent state-projection layer, we first find that this standard validation certifies a suite in which a first-order mutant removing event-identity deduplication survives every check. We then freeze the repaired twelve-check suite, record its hash, and run it once against ten mutants specified by an adversarial reader who designed none of its fixtures: it kills five. Instrumenting the five survivors against the reference shows they fail in \emph{two} distinct ways, not one. Three are never activated, because no fixture supplies an input on which the mutated code behaves differently at all. The other two corrupt internal state that no oracle in the suite can observe. The two modes need different repairs, and neither is visible from a pass/fail report. Treating the missing inputs as a coverage question, we enumerate seven discriminating dimensions of the input space, register in advance which are uncovered and which survivors they should explain, and add one fixture per uncovered dimension while reusing the existing oracles verbatim. All five survivors then die, each to the check written for its predicted dimension. We report this as a repair result on the same challenge set rather than a second held-out estimate, and give the artifact, including the frozen hash, the registered predictions, all mutants and the run logs, so the distinction is checkable.
\end{abstract}

\section{Introduction}
\label{sec:intro}

Agentic systems that operate over multiple turns and multiple concurrent sessions are typically wrapped in a layer of automated checking: invariant assertions, runtime monitors, safety validators, or judge models that decide whether an observed trajectory is acceptable. These checks are then used, directly or indirectly, to gate deployment, trigger human escalation, or certify that a system behaves as intended. The evidentiary weight placed on this layer is substantial, and yet the layer itself is rarely subjected to the same scrutiny as the system it monitors. A verification suite that never triggers a failure is observationally identical, from outside the suite, to a system that never has one.

\citet{reuel2025openproblems} name this gap as an open problem in technical AI governance: it is hard to be confident that an evaluation's results reflect the property being measured rather than the evaluation methodology itself. They pose it for models. We ask it about the infrastructure agents run on. The technique for answering it at the software layer is mutation analysis \citep{demillo1978hints,jia2011analysis}, and the conditions under which a fault escapes are standard \citep{ammann2016introduction}: it must be reached, must infect state, and that infection must propagate to an observed output. We take the technique as given. The question we bring to it is not the one mutation analysis usually answers. Mutation score is normally a developer-facing measure of test adequacy, consumed to decide where to write more tests. Here the suite's output is offered as \emph{evidence}, to license a deployment or an intervention, so the operative question is what a verifier's silence entitles anyone to conclude. That question has an answer we can measure, and the answer turns out to be governed by something other than the properties the suite asserts.

We study the client-side \emph{projection} layer of a multi-session agent chat interface, which rebuilds visible conversation state from asynchronous events (WebSocket pushes, HTTP acknowledgments, periodic snapshots) arriving unreliably, per session, per user. Six failure classes follow from asynchronous multi-channel replication: duplicate delivery, non-idempotent replay, cross-session leakage, out-of-order rendering, non-convergent concurrent updates, and stale-state regression. We write a reference implementation handling all six, an eleven-check invariant suite, and first-order mutants each altering exactly one mechanism.

The paper is one argument in three steps. \textbf{The validation everyone uses is uninformative.} Passing the reference and failing an all-defects variant certifies our suite while a first-order mutant survives all eleven checks, because the reference enforces the property twice and no fixture separates the two enforcements. \textbf{Measured against externally specified mutants, the certified suite resolves half.} We froze the repaired twelve-check suite, recorded its hash, and ran it once against ten mutants specified by an adversarial reader of this manuscript who took no part in designing any fixture: five survive. \textbf{The survivors fail in two distinct ways, and the distinction dictates the repair.} Instrumenting each survivor against the reference at every step of every fixture shows that three are never activated. The mutated code runs but produces identical internal state, because no fixture supplies an input on which the change matters. Two others do corrupt internal state that no oracle in the suite can observe. Both are coverage problems, but of different kinds: the first needs new input \emph{values}, the second needs input values that carry an existing corruption to an observable output. We enumerate seven dimensions along which the reference's control flow can discriminate two executions, register which are uncovered before writing anything, and add one fixture per uncovered dimension while reusing the existing oracle predicates unchanged. All five survivors then die, each to the check written for its predicted dimension.

\paragraph{Contributions.}
\begin{enumerate}[topsep=2pt,itemsep=1pt]
    \item \textbf{A measurement where there was previously an assurance.} We report a post-freeze mutation kill rate for a verification suite guarding agent infrastructure: \textbf{5 of 10} externally specified first-order mutants, with the frozen-suite hash, the mutants, and the survivors published. The usual evidence for such a suite is an unbroken record of not firing, which is also what a suite of no resolution produces. (The 10/10 reported after repair is a result on the same challenge set, not a second held-out estimate, and we keep the two separate throughout.)
    \item \textbf{A demonstration that the standard validation cannot detect this.} The correct/all-defects pair certifies a suite that a first-order mutant survives outright. The mechanism is redundancy in the system under test: two guards independently suppress a redelivered event, so no fixture exercising redelivery can attribute the outcome to either. This is the masking problem that condition-coverage criteria such as MC/DC exist to address, arising here between two implementation mechanisms rather than two clauses of one predicate.
    \item \textbf{A two-mode account of why a suite misses, measured rather than asserted.} We instrument every mutant against the reference at each step of each fixture and record where the two first diverge. Three survivors never diverge internally at all. Two diverge internally but never visibly. A pass/fail report cannot distinguish these, and they call for different repairs. We are not aware of this breakdown being reported for a verification suite used as deployment evidence.
    \item \textbf{A registered prediction that the repair confirms.} Treating the gap as an input-space coverage problem, we enumerate seven discriminating dimensions, record in advance which are uncovered and which survivors they should account for, then add one fixture per uncovered dimension with the oracles reused verbatim. All five predictions hold.
    \item \textbf{A negative result on invariant-family taxonomies.} Mutants co-designed with the suite's five families localise cleanly to their own family. Externally specified mutants do not: a deduplication fault is caught by a convergence check, and two unrelated snapshot faults by the same check. Families organise checks well and describe coverage badly.
\end{enumerate}

\section{Related Work}
\label{sec:related}

\paragraph{Contract- and invariant-based verification of agent infrastructure.}
The closest work to ours is \citet{paduraru2026trace}, who propose a trace-based assurance framework for agentic orchestration combining contracts, testing, and governance. Their framework validates itself through fault injection and budgeted counterexample search over an orchestration graph, targeting failures such as non-termination, role drift, and claim propagation across agents. Our target failure surface is different, namely client-side session-state consistency rather than orchestration semantics, and more importantly our validity argument is different in kind: rather than searching for counterexamples against a specification, we construct a fixed, interface-compatible mutant known in advance to violate the specification, and ask whether the checking apparatus notices. The two approaches are complementary: fault injection is well suited to open-ended search for unknown failure modes, while a differential mutation argument is well suited to auditing whether \emph{known} failure modes are reliably caught.

\paragraph{Empirical failure taxonomies for deployed agents.}
\citet{wu2026silent} give a longitudinal taxonomy of silent failures in a production LLM agent runtime, drawn from postmortems. That work characterises what goes wrong. Ours asks whether an automated checker would have caught it.

\paragraph{Active testing of agent frameworks.}
\citet{long2026logichunter} test popular agent frameworks using fuzzing combined with an active, ReAct-style oracle that inspects runtime state during execution. The oracle mechanism there is active inspection against a library's exposed API. Ours is passive differential replay against a fixed set of scenarios, and the target is a deployed session-state layer rather than a framework itself.

\paragraph{Evaluation of monitors rather than systems.}
\citet{guerin2022unifying} evaluate the discriminative power of machine-learning safety monitors using detection metrics against labeled data. We have no labeled corpus of real failures and construct the fault set by design instead, which is what makes a kill rate available at all.

\paragraph{Relation to mutation testing.} Constructing faulty program variants to measure whether a test suite detects them is mutation analysis, with a mature literature on first- versus higher-order mutants, surviving mutants as evidence of test inadequacy, mutation score as an adequacy criterion, and equivalent-mutant detection \citep{demillo1978hints,jia2011analysis}. Our five single-fault variants are first-order mutants and the all-defects variant is a higher-order mutant. A suite that kills the higher-order mutant while a constituent first-order mutant survives is exactly the situation mutation analysis exists to expose. Our contribution is a post-freeze mutation score for a verification suite guarding agent infrastructure, together with a per-mutant account of where in the reachability--infection--propagation chain each surviving fault stops. We are not aware of either being reported for a checking layer used as deployment evidence. \citet{just2014mutants} show that mutant detection correlates with real-fault detection, which is what makes a mutation score meaningful as evidence about a suite.

\paragraph{Relation to metamorphic and differential testing.} Redelivery, replay, reordering, and convergence under permuted arrival are metamorphic relations between executions rather than assertions about a single output, and the survivors we report are failures of source-test selection under those relations \citep{segura2016survey}. Differential testing classically compares independently written implementations as pseudo-oracles, and has a long history for stateful and distributed systems \citep{statediver2022,acto2023}. We do not do that. Our oracle is explicit invariant assertions, and the distinction matters for what our result can claim.

\section{The Verification Target: A Multi-Session Projection Layer}
\label{sec:target}

We study a common architectural pattern in multi-session agent chat interfaces: each session maintains a client-side \emph{projection}, a visible message list and a run-status field, built by applying a stream of asynchronous events to a prior state. Events arrive through more than one channel: an optimistic local echo when the user sends a message, an HTTP acknowledgment once the backend accepts it, a WebSocket push once the agent responds, and a periodic snapshot for reconnection and late-joining clients. We formalize the interface any implementation of this layer must satisfy as a \texttt{ProjectionAdapter}: given a session identifier it produces an initial state, given a state and an event envelope it produces a new state, and it can render the visible messages from any state. Three additional hooks are optional, \texttt{applyOptimistic}, \texttt{applyHttpAck} and \texttt{applySnapshot}, and a check that needs one skips (rather than silently passing) when an adapter does not implement it.

This pattern admits six failure classes, which motivate the checks in Section~\ref{sec:method}:

\begin{enumerate}[topsep=2pt,itemsep=1pt,leftmargin=*]
    \item \textbf{Non-idempotent delivery.} The same event, delivered twice (e.g., after a network retry), changes the visible projection rather than leaving it unchanged.
    \item \textbf{Non-idempotent replay.} Re-consuming the full event stream from the start (e.g., on a reconnect that lacks a resumption cursor) changes the projection relative to consuming it once.
    \item \textbf{Cross-session leakage.} An event addressed to one session is applied to another, typically the result of a broadcast channel that is not scoped by session identifier before dispatch.
    \item \textbf{Order-insensitive rendering.} Events that arrive out of logical order are rendered in arrival order, with no buffering, gap detection, or explicit resynchronization signal.
    \item \textbf{Non-convergent concurrent update paths.} The same logical message reaches the projection through more than one channel, an optimistic echo, an HTTP acknowledgment and a WebSocket push, and the three paths are reconciled differently depending on arrival order, rather than converging to one canonical message regardless of order.
    \item \textbf{Stale-state regression.} A late-arriving snapshot or status update overwrites a projection that has already advanced past it, silently regressing terminal or otherwise-settled state.
\end{enumerate}

We take these six classes as given, as recurring consequences of this architectural pattern, and use them to construct the seven adapters described next. Nothing about the six classes or the adapters below is a claim about any specific deployed system. They are a fixed target for validating a checking methodology, not an audit of one.

\section{Method: A Reference Implementation and Its Mutants}
\label{sec:method}

\paragraph{Reference adapter.} The correct implementation deduplicates by event identifier, upserts by message identifier rather than appending, scopes every event to its originating session before mutating state, buffers out-of-order events behind a monotonic cursor and flags a gap rather than rendering past it, reconciles the optimistic/acknowledgment/WebSocket paths for the same logical message into one entry regardless of arrival order, merges snapshots into live state rather than replacing it, and treats run-status updates as monotonic (a terminal status cannot be overwritten by a stale in-flight one).

\paragraph{All-defects mutant.} The all-defects mutant, \naiveadapter{}, is constructed, not discovered: it appends on every event with no deduplication or upsert logic, applies no session-scope guard before mutating state, renders strictly in arrival order with no buffering or gap detection, lets the HTTP-acknowledgment and WebSocket paths both append independently without reconciliation, replaces the entire projection on every snapshot rather than merging it, and takes the last write on run-status regardless of whether it regresses a terminal state. Each of these six choices is a direct, by-construction instantiation of one failure class from Section~\ref{sec:target}. Its faults are built in rather than inferred from observed behaviour, so whether a check catches them is a question about the check and not about ground truth.

\paragraph{First-order mutants.} A correct/all-defects pair separates two extremes, which a suite of very low resolution also does: firing on everything passes the broken side, firing on nothing passes the correct side. We therefore construct five first-order mutants, each identical to \refadapter{} except for one injected fault, one per invariant family, a choice whose effect on family localisation Section~\ref{sec:results} measures.

\begin{itemize}[topsep=2pt,itemsep=0pt,leftmargin=*]
  \item \dupdefect{}: the event-identity deduplication step is removed. Redelivered events are reapplied.
  \item \regressdefect{}: the version guard on run-status updates is removed. Last write wins, so a stale replay can regress a terminal status.
  \item \scopedefect{}: the session-scope guard is removed. An event addressed to another session mutates this one.
  \item \partialadapter{}: on a gap, the cursor advances to the highest sequence \emph{observed} rather than the highest contiguous sequence \emph{applied}. This is an ordinary mistake, tracking ``latest seen'' instead of ``last contiguous'', and it compounds: once the cursor jumps the gap, the missing event is later rejected as stale, so the gap never heals.
  \item \snapdefect{}: snapshots replace the projection instead of merging into it, so a late snapshot erases live state.
\end{itemize}

Every other line of each mutant is the reference implementation verbatim, so a firing check is responding to one syntactic change. It does not follow that the change has one behavioural consequence: a cursor or merge fault can violate several properties at once, and Section~\ref{sec:results} shows that it does.

\paragraph{Invariant suite.} We start from eleven checks (\texttt{C-01}--\texttt{C-11}), grouped into five invariant families, exercised against eleven fixed, deterministic scenarios (duplicate delivery, full-stream replay, distinct-but-similar messages, a foreign-session event, a sequence gap, out-of-order delivery, three orderings of optimistic/ack/WebSocket convergence, two orderings of snapshot-plus-live-events convergence, a stale status update, and a same-sequence conflict). The five families are: \textsc{idempotence} (duplicate and full-stream replay must not change the projection), \textsc{distinctness / no regression} (messages that should stay distinct must not be merged, and terminal state must not regress), \textsc{isolation} (a session must not be mutated by another session's event), \textsc{ordering} (gaps must not silently advance the cursor, out-of-order delivery must render correctly or flag itself, and duplicate-sequence conflicts must not both silently win), and \textsc{convergence} (all update paths for one logical message, and snapshot-plus-live-events under either arrival order, must reach the same state). Section~\ref{sec:results} adds a twelfth check, then five more, for reasons the mutants supply.

\begin{figure}[t]
\centering
\includegraphics{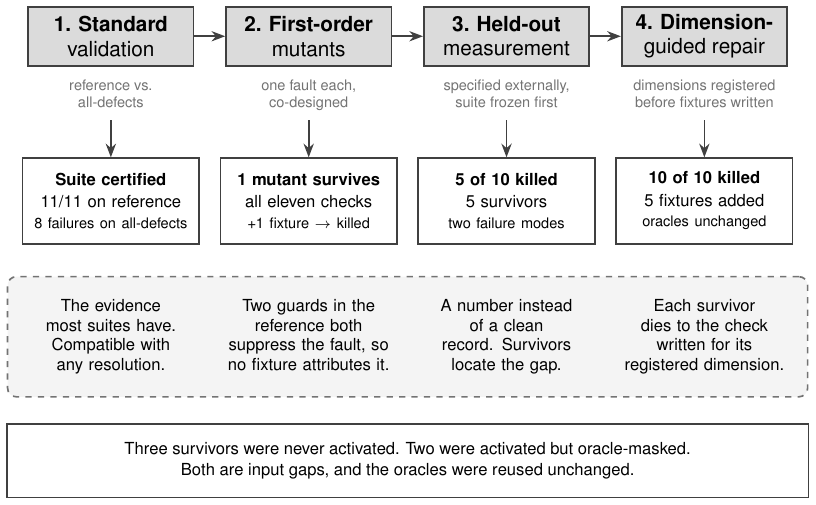}
\caption{Four stages of evidence about one suite. Stage 1 is the validation such a suite normally
receives, and it certifies this one. Stage 2 shows what the certificate omits: a mutant removing a
single guard survives every check, because the reference enforces the property twice and no fixture
separates the two enforcements. Stage 3 is the measurement, against mutants specified by a reader
who designed none of the fixtures, on a suite frozen and hashed beforehand. Stage 4 is a repair
evaluated on those same survivors, not a second measurement: its value is that the dimensions and
their per-survivor predictions were registered before any fixture was written.}
\label{fig:construction}
\end{figure}

\paragraph{Requirements on the suite.} Figure~\ref{fig:construction} summarizes the four stages. Two requirements are uncontroversial: the suite must not fire on \refadapter{}, and it must fire on \naiveadapter{}, since a suite that never fires certifies nothing. Both are satisfied, and Section~\ref{sec:results} shows that satisfying both is compatible with carrying a hole. The requirement that does discriminate is a post-freeze one: given mutants specified by someone who did not design the fixtures, against a suite frozen before those mutants exist, what fraction does it kill? Family localisation, where a single-fault mutant fails only inside its own invariant family, is not among the requirements: Section~\ref{sec:results} shows it is a property of how mutants are chosen rather than of the suite.

\section{Results}
\label{sec:results}

\subsection{Outcome of the Standard Validation}

Against the reference/all-defects pair the eleven-check suite behaves as a validated instrument should: \refadapter{} passes all eleven, \naiveadapter{} fails eight and is skipped on one. This is the evidence such a pair can produce, and it is the evidence most suites have.

A first-order mutant that removes event-identity deduplication and changes nothing else fails \emph{zero} of the eleven, including both checks whose stated subject is idempotence under redelivery. Two properties of the fixture set independently prevent the mutant from becoming distinguishable from the reference. The duplicate and replay fixtures deliver both copies with a sequence number, and the reference rejects any event whose sequence does not exceed the cursor, so the ordering guard suppresses the redelivery before the absent deduplication matters. Independently, those fixtures redeliver \texttt{message.created}, applied by upserting on \texttt{messageId}. Writing the same key twice yields the same state, so even unsequenced the duplicate leaves no trace. The idempotence family was not testing deduplication. It was testing ordering.

The repair is a check-level change even though the cause is a fixture: an input that is simultaneously \emph{unsequenced} and \emph{append-shaped} defeats both suppressors. We add one, a streaming \texttt{message.delta} with a repeated \texttt{eventId} and no sequence, as C-12.

\subsection{Post-Freeze Mutation Kill Rate}

Repairing against the mutant that exposed a hole and then reporting that the mutant is caught measures nothing. We froze the twelve-check suite, recorded its hash, and only then obtained ten further first-order mutants specified by an adversarial reader of this manuscript who had the system description and the check titles but designed none of the fixtures and did not see their contents. All ten specifications were implemented, and none was discarded as invalid or equivalent. The suite was run once and not modified.

\textbf{It killed five of ten} (Table~\ref{tab:heldout}, column A). \refadapter{} still passed 12/12, so the survivors are not an artifact of a suite that had stopped firing. Because the mutants were specified adversarially rather than sampled, 5/10 is a challenge-set score and not an estimate of the detection rate under any natural fault distribution.

\textbf{Detection across invariant families.} The five mutants written alongside the families each fail only inside their own family. The externally specified set does not. H1 keys deduplication on \texttt{messageId} rather than \texttt{eventId}, an idempotence fault under any reasonable taxonomy, and is caught only by C-08, a \emph{convergence} check, because the optimistic/acknowledgment/WebSocket path is the sole fixture where the two keys diverge. H4 and H9 are structurally unrelated snapshot faults and are both caught by C-09 alone. Family localisation is a property of how mutants are chosen, not of the suite.

\subsection{Two Modes of Missing a Fault}

Why a suite misses is not visible from its pass/fail output, so we measured it. For every (mutant, fixture) pair we replayed the fixture step by step against both the reference and the mutant, serialising the full internal state and the observable projection after each step, and recorded where the two first diverge. A fault that never diverges internally was never activated by that input. One that diverges internally but never observably is masked at the oracle boundary. Only the latter is in reach of a better assertion on the same inputs.

Under the twelve pre-repair fixtures, seven of the ten mutants are activated and five of those propagate to an observable output. The five that propagate (H1, H4, H5, H7 and H9) are exactly the five the suite kills, and the five survivors split into the two modes below.

\begin{itemize}[topsep=2pt,itemsep=0pt,leftmargin=*]
  \item \textbf{Never activated (H6, H8, H10).} The mutated code executes but computes the same result. H6 compares session identifiers case-insensitively and every fixture uses identifiers differing by more than case, so the altered comparison returns what the original returned. H8 permits an equal-version regression and the stale-status fixture supplies a strictly lower version, so the widened branch is never taken. H10 lets an acknowledgment overwrite a richer status and no fixture delivers an acknowledgment after the event it acknowledges. No assertion over these inputs could have caught any of the three, however strong.
  \item \textbf{Activated but oracle-masked (H2, H3).} Internal state does diverge. H2 removes deduplication for non-message events, so the mutant's seen-event set differs from the reference's after the first \texttt{run.*} event, and no fixture then redelivers one, so the divergence stays internal. H3 bounds the deduplication cache to a single entry, and the difference is real from the second distinct event onward, but no fixture redelivers an event after an intervening one. These two are within reach of a wider oracle as well as of new inputs.
\end{itemize}

The distinction matters because it separates two repairs that a bare kill rate conflates. Three of five gaps could only ever be closed by new input values. Two could be closed either by new inputs or by exposing more internal state to the checks. We take the input route for both, because it keeps the oracles fixed and therefore keeps the experiment interpretable, but a practitioner facing H2 or H3 has a genuine choice.

\subsection{Discriminating Fixture Coverage as a Registered Prediction}

If the survivors fail because no input distinguishes mutant from reference, then what a suite resolves is fixed by the shapes of its inputs, and input shape is enumerable in a way that ``assert more properties'' is not. This is testable. Before writing any new fixture we enumerated the dimensions along which the reference's control flow can discriminate two executions, recorded which the existing twelve fixtures cover, and recorded which survivors the uncovered ones should account for.

\begin{samepage}
\begin{itemize}[topsep=2pt,itemsep=0pt,leftmargin=*]
  \item[\textbf{D1}] event shape: upsert-shaped vs.\ append-shaped \hfill \emph{covered}
  \item[\textbf{D2}] sequencing: sequenced vs.\ unsequenced \hfill \emph{covered (by C-12)}
  \item[\textbf{D3}] event family: \texttt{message.*} vs.\ \texttt{run.*} \hfill \textbf{uncovered $\rightarrow$ H2}
  \item[\textbf{D4}] redelivery distance: immediate vs.\ after an intervening event \hfill \textbf{uncovered $\rightarrow$ H3}
  \item[\textbf{D5}] session identity: exact vs.\ differing only in case \hfill \textbf{uncovered $\rightarrow$ H6}
  \item[\textbf{D6}] version relation: strictly lower vs.\ equal \hfill \textbf{uncovered $\rightarrow$ H8}
  \item[\textbf{D7}] channel order: WebSocket before ack vs.\ ack after \hfill \textbf{uncovered $\rightarrow$ H10}
\end{itemize}
\end{samepage}

We generated five fixtures, one per uncovered cell, without consulting any mutant's source, and added checks C-13 to C-17 that reuse existing oracle predicates.

\textbf{All five survivors died} (Table~\ref{tab:heldout}, column B), and every registered prediction held: H2 to C-13, H3 to C-14, H6 to C-15, H8 to C-16, H10 to C-17. One correspondence is not exclusive. C-13, written for the \texttt{run.*} dimension, also kills H3, which is worth reporting because it shows the dimensions are not orthogonal: a redelivered \texttt{run.*} event also sits at distance one from its original. \refadapter{} passes all seventeen checks, so the added fixtures produce no failures on correct code.

\textbf{Repair result versus held-out measurement.} The dimensions were derived by inspecting the same five survivors the repair is evaluated on. What the experiment establishes is that a written-down diagnosis correctly predicted which inputs would expose which faults, five times out of five, before those inputs existed. What it does not establish is that the seventeen-check suite would score 10/10 against unseen mutants, or that seven dimensions are complete.

\paragraph{Oracle reuse.} Each new check pairs a new input with an assertion already in the suite (Table~\ref{tab:oracles}). This is what licenses locating the gap in the inputs: had we also strengthened the assertions, the two causes would be confounded.

\begin{table}[t]
\caption{Externally specified mutants against the frozen suite. \ding{51}\ means the mutant was \emph{killed} (at least one check failed). \ding{55}\ means it survived. Column A: the twelve-check suite, frozen and hashed before any of these mutants was written. Column B: after adding one fixture per uncovered dimension, oracles unchanged. The dimension column records the prediction registered before the fixtures existed. \refadapter{} fails no check in either column (12/12 and 17/17).}
\label{tab:heldout}
\centering
\small
\begin{tabularx}{\textwidth}{@{} l X c c l @{}}
\toprule
& Injected fault & A & B & Killed by (registered dim.) \\
\midrule
H1  & deduplication keyed on \texttt{messageId}       & \ding{51} & \ding{51} & C-08 \emph{(convergence)} \\
H2  & deduplication only for \texttt{message.*}       & \ding{55} & \ding{51} & C-13 \quad(D3) \\
H3  & deduplication cache bounded to one entry        & \ding{55} & \ding{51} & C-14, also C-13 \quad(D4) \\
H4  & snapshot resets the cursor unconditionally      & \ding{51} & \ding{51} & C-09 \\
H5  & over-aggressive merge on role\,+\,text          & \ding{51} & \ding{51} & C-03 \\
H6  & session identifier compared case-insensitively  & \ding{55} & \ding{51} & C-15 \quad(D5) \\
H7  & off-by-one contiguity test                      & \ding{51} & \ding{51} & C-05, C-06 \\
H8  & equal-version update regresses a terminal state & \ding{55} & \ding{51} & C-16 \quad(D6) \\
H9  & snapshot drops live-only messages from order    & \ding{51} & \ding{51} & C-09 \\
H10 & acknowledgment overwrites a richer status       & \ding{55} & \ding{51} & C-17 \quad(D7) \\
\midrule
\multicolumn{2}{@{}l}{\textbf{Killed}} & \textbf{5/10} & \textbf{10/10} & \\
\bottomrule
\end{tabularx}
\end{table}

\begin{table}[t]
\caption{The five added checks reuse existing oracle predicates unchanged, and only the input differs. \texttt{digestOf} compares the rendered projection before and after a step, \texttt{runStatus} reads a run's status, \texttt{visibleMessages} reads the rendered list.}
\label{tab:oracles}
\centering
\small
\begin{tabularx}{\textwidth}{@{} l X l @{}}
\toprule
New check & New input dimension & Oracle predicate, reused from \\
\midrule
C-13 & \texttt{run.*} event family (D3)               & \texttt{digestOf} + \texttt{runStatus}, as C-01 and C-10 \\
C-14 & redelivery after an intervening event (D4)     & \texttt{digestOf} before/after, as C-01 \\
C-15 & session identifier differing only in case (D5) & \texttt{digestOf} before/after, as C-04 \\
C-16 & equal version rather than strictly lower (D6)  & \texttt{runStatus}, as C-10 \\
C-17 & acknowledgment after the WebSocket event (D7)  & \texttt{digestOf} + \texttt{visibleMessages}, as C-07 \\
\bottomrule
\end{tabularx}
\end{table}

\section{Discussion}
\label{sec:discussion}

\paragraph{Activation gaps and masking gaps.} The usual response to an inadequate suite is to assert more properties. For three of our five survivors that response could not have worked: the mutated code produced identical internal state on every fixture, so no assertion over those inputs had anything to distinguish. For the other two an assertion over internal state would have worked, and so did a new input. Distinguishing the modes requires instrumenting mutant against reference, which a pass/fail report does not do, and it changes what the practitioner should do next. This is also why we take the input route uniformly: it holds the oracles fixed, so the repair is attributable.

\paragraph{Evidential value of a monitor's silence.} A suite that has never fired unexpectedly is compatible with a suite that cannot fire. Both produce the same record, and that record is what is usually offered when a monitor's output licenses a deployment. A post-freeze challenge score breaks the tie: it is falsifiable, degrades visibly when the suite is weakened, and can be recomputed after any change. Ours went from certified-by-the-standard-method to 5/10 against an adversarial challenge set, a statement about what a determined reader could get past this suite.

\paragraph{Redundancy in the system under test.} The original miss was not a coding error. Every check does what its title says. Deduplication by event identity and rejection by sequence staleness both suppress a redelivered event, so no fixture exercising redelivery can attribute the outcome to either. Condition-coverage criteria such as MC/DC exist precisely to force inputs on which one condition independently determines the outcome \citep{hayhurst2001mcdc}. The same masking arises here between two implementation mechanisms rather than two clauses of one predicate, and it is not addressed by any coverage criterion the suite tracks.

\paragraph{Relation to input-space partitioning.} Our seven dimensions are a small input-parameter model in the tradition of category-partition and combinatorial testing \citep{ammann2016introduction,kuhn2013combinatorial}. What is new here is the direction of use: rather than deriving partitions from a specification up front, we use externally specified surviving mutants to reveal which partitions a verification suite lacks, and then test whether the resulting model predicts the repair. The model extends as challenge sets grow: redelivery distance beyond one intervening event (a cache bounded to two entries needs distance two), session identity under whitespace, Unicode normalisation and encoding variants, version relations with absence and wraparound, and channel orders over longer traces.

\paragraph{Invariant families as an organising unit.} Grouping checks into families is useful for writing and reading them, and is the natural unit in which coverage gets described. Our data show it is the wrong unit for the second purpose: a deduplication fault surfaced in a convergence check, and two unrelated snapshot faults in the same check. Coverage arguments of the form ``every family has checks'' are therefore not arguments about detection.

\paragraph{Deployment semantics of a skip.} One check cannot be attempted against \naiveadapter{}, which exposes neither hook it requires. Three-valued verdicts are standard in runtime verification. What a gate additionally needs is the decision attached to the third value: for a property whose violation is unrecoverable, inability to observe it plausibly has to fail closed.

\paragraph{Scope.} All implementations are synthetic and deterministic. The controlled design establishes internal validity of the mutation outcomes, since we know exactly which fault is present, but limits external validity to hand-written first-order faults, which are an imperfect proxy for those developers introduce \citep{just2014mutants}. The reader who specified the held-out mutants saw the system description and the check titles but not the fixtures, which makes the set adversarial with respect to the stated properties rather than blind. Two extensions follow directly: a second challenge round against the frozen seventeen-check suite by a reader who has not seen these survivors, and the same checks bound to an implementation the authors did not write.

\section{Conclusion}
\label{sec:conclusion}

The evidence normally offered for a verification suite guarding agent infrastructure is that it passes code believed correct and fails code believed broken. We measured what that evidence is worth for one such suite and found it worth little: the certified suite was survived outright by a first-order mutant, and killed five of ten faults specified by a reader who had not designed its fixtures.

The failures were not scattered, but neither were they uniform. Instrumenting each survivor against the reference showed two modes: three faults that no fixture ever activated, and two that corrupted internal state no oracle could observe. Only the second kind is reachable by a stronger assertion on the same inputs, so a bare kill rate conflates two problems with different remedies. Both are gaps in the input set, and treating them that way is testable: we enumerated the dimensions along which the implementation can discriminate two executions, registered which were uncovered and which survivors they should explain, and added one fixture per uncovered dimension with the oracles unchanged. All five predictions held, and because they were registered before any fixture existed, the repair is checkable as a prediction on the survivors that motivated it rather than read as a second measurement.

For a layer whose output is meant to license a deployment decision, the failure direction that matters is not a monitor that fires too readily but one that stays quiet in a way a passing report presents as success. The procedure that surfaces this needs little beyond what such a suite already has: each mutant is the reference implementation with a single edit, and the measurement is one run of the existing suite.

\bibliographystyle{plainnat}

\end{document}